\documentclass[apj]{emulateapj}
\begin{document}
\title{The Compact X-ray Source 1E 1547.0-5408 and the Radio Shell
  G327.24-0.13: A New Proposed Association between a Candidate Magnetar
  and a Candidate Supernova Remnant}
\author{Joseph D. Gelfand}
\affil{Harvard-Smithsonian Center for Astrophysics} 
\affil{60 Garden Street Cambridge, MA 02138} 
\email{jgelfand@cfa.harvard.edu} 
\and
\author{B. M. Gaensler\altaffilmark{1}} 
\affil{Harvard-Smithsonian Center for Astrophysics, Cambridge, MA
  02138} 
\and 
\affil{School of Physics A29, The University of Sydney, NSW 2006, Australia}
\email{bgaensler@usyd.edu.au}

\altaffiltext{1}{Alfred P. Sloan Research Fellow,
  Australian Research Council Federation Fellow}

\begin{abstract}
We present X-ray, infrared and radio observations of the field
centered on X-ray source 1E~1547.0--5408 in the Galactic Plane.
Analysis of a new {\it Chandra} observation of this source shows it is
unresolved at arc-second resolution, and analysis of a new {\it XMM}
observation shows that its X-ray spectrum is best described by an
absorbed power-law and blackbody model.  A comparison of the X-ray
flux observed from 1E~1547.0--5408 between 1980 and 2006 by {\it
Einstein}, {\it ASCA}, {\it XMM}, and {\it Chandra} reveals that its
absorbed 0.5--10~keV X-ray flux decreased significantly during this
period, from $\sim2\times10^{-12}$~ergs~cm$^{-2}$~s$^{-1}$ to
$\sim3\times10^{-13}$~ergs~cm$^{-2}$.  No pulsations in the X-ray
emission from 1E~1547.0--5408 were observed during the most recent
{\it XMM} observation, which allows us to put a 5$\sigma$ confidence
upper limit of 14\% for the 0.5--10~keV peak-to-peak pulsed fraction
(for sinusoidal pulses with periods slower than 1.8s).  A
near-infrared observation of this field shows a source with magnitude
$K_s = 15.9\pm0.2$ near the position of 1E~1547.0--5408, but the
implied X-ray to infrared flux ratio indicates the infrared emission
is most likely from an unrelated field source, allowing us to limit
the IR magnitude of any counterpart to 1E~1547.0--5408 to $\ga
17.5$. Archival radio observations reveal that 1E~1547.0--5408 sits at
the center of a faint, small ($4\arcmin$ diameter) radio shell,
G327.24--0.13, which is possibly a previously unidentified supernova
remnant.  The X-ray properties of 1E~1547.0--5408 suggest that this
source is a magnetar --- a young neutron star whose X-ray emission is
powered by the decay of its extremely strong magnetic field, $B \sim
10^{14-15}$~G.  The spatial coincidence between this source and
G327.24--0.13 suggests that 1E~1547.0--5408 is associated with a young
supernova remnant, supporting a neutron star interpretation.
Additional observations are needed to confirm the nature of both
1E~1547.0--5408 and G327.24--0.13, and to determine whether these
sources are physically associated.  If confirmed, this pair will be an
important addition to the small number of known associations between
magnetars and supernova remnants.
\end{abstract}

\section{Introduction}
\label{intro}
Massive stars ($M \ga 8 M_{\odot}$) end their lives in supernovae,
often forming neutron stars.  For many decades, it was thought that
these neutron stars had short initial spin periods ($P_0 \ll 1~{\rm
s}$), dipole surface magnetic fields with strengths $B \sim
10^{12}~{\rm G}$, and were most often observed as radio pulsars.
While this describes the majority of known neutron stars, several new
classes of neutron stars have since been discovered.  Most notable of
these are Anomalous X-ray Pulsars (AXPs) and Soft Gamma Repeaters
(SGRs), which have spin periods longer than most normal radio pulsars
($P\sim5-12$~s) and high period-derivatives ($\dot{P}\sim10^{-11}$
s/s) \citep{woods04}.  Due to the strong surface dipole magnetic
fields inferred from these timing properties ($B \sim 10^{14-15}~{\rm
G}$), these sources are believed to be ``magnetars'', neutron stars
whose X-ray emission is powered by the decay of these extremely strong
field \citep{duncan92,thompson95,thompson96}.

Currently there are only 13 confirmed magnetars.  As a result, it is
difficult to determine the spatial distribution, birth-rate, and
active lifetime of these sources, information vital to understanding
the relationship between magnetars and other products of core collapse
supernovae.  This uncertainty is further exacerbated by the presence
of at least one transient magnetar, XTE~J1810-197 \citep{ibrahim04}.
A powerful way of determining the relationship between magnetars and
other classes of neutron stars is to use environmental information to
constrain their ages and progenitors of these populations.  This is
easiest done for those neutron stars associated with SNRs, since
observations of SNRs allow an independent estimates of distances and
ages, the densities of the surrounding medium, and the explosion
energies of the progenitor supernovae.  To date, there are only two
secure magnetar/SNR associations, 1E~2259+586 in SNR CTB 109
\citep{fahlman81} and 1E~1841-045 in SNR Kes~73 \citep{vasisht97}, 
though SGR~0526-66 is possibly associated with Large Magellanic Cloud
SNR N49 \citep{marsden96,gaensler01}.  Based on the ages of these SNRs
(not including SNR~N49), as well as the offset between these magnetars
and the centers of their SNRs, \citet{gaensler01} concluded that
magnetars are young ($<10000$~yr) neutron stars with low projected
space velocities ($<$500~km~s$^{-1}$).  This assertion that magnetars
have a low spatial velocity is also supported by recent proper motion
measurements of XTE~J1810-197 \citep{helfand07}.  Additionally, an
analysis of the X-ray emission from these SNRs implies that the level
of energy injection at early times, from both supernova ejecta and
from the magnetar spin down, was $\sim10^{51}$~ergs, constraining the
initial spin period of these magnetars to $P_0 \ga 5$~ms
\citep{vink06}.  While much has been learned from detailed studies of
the few associations between magnetars and SNRs, much more can be
gained by identifying additional examples of such associations.

In this paper, we propose a new potential magnetar/SNR association,
between the X-ray source 1E~1547.0-5408 and the Galactic radio shell
G327.24-0.13.  In \S\ref{1e1547}, we present both X-ray (\S\ref{xray})
and near-infrared (near-IR) (\S\ref{nir}) observations of the field
around 1E~1547.0-5408.  In \S\ref{radio}, we present archival radio
observations which show that 1E~1547.0-5408 is located at the center
of G327.24-0.13.  In \S\ref{conclusions}, we argue the compact nature
of 1E~1547.0-5408, its lack of a bright near-IR counterpart, and its
location of 1E~1547.0-5408 source at the center of G327.24-0.13
implies this source is a neutron star, and that the X-ray spectrum and
variability of 1E~1547.0-5408 strongly suggest that it is a magnetar.

\section{Observations of 1E~1547.0-5408}
\label{1e1547}

\subsection{X-ray Data}
\label{xray}

X-ray source 1E~1547.0-5408 was discovered by {\it Einstein} during a
search for X-ray counterparts of unidentified $\gamma$-ray sources
\citep{lamb81}, in this case the $\gamma$-ray source 1CG327--0
detected by {\it Cos-B} \citep{hermsen77}.  This detection was
confirmed by the {\it ASCA} Galactic plane survey, in which the source
AX~J155052-5418 was detected at the same position \citep{sugizaki01}.
Recently, this field has been observed twice with {\it XMM--Newton},
(an archival observation in 2004 and a new observation in 2006), and
again with the {\it Chandra X-ray Observatory} in 2006.  A summary of
these observations is given in Table \ref{obssum}.  Both {\it XMM}
datasets were processed using the standard tasks given in {\sc
xmm-sas~v.6.5.0}, while the {\it Chandra} data were analyzed using
{\sc ciao~v.3.4}.  All datasets were then filtered using the standard
energy and quality criteria.

\begin{figure}
\begin{center}
\includegraphics[angle=0,scale=0.45]{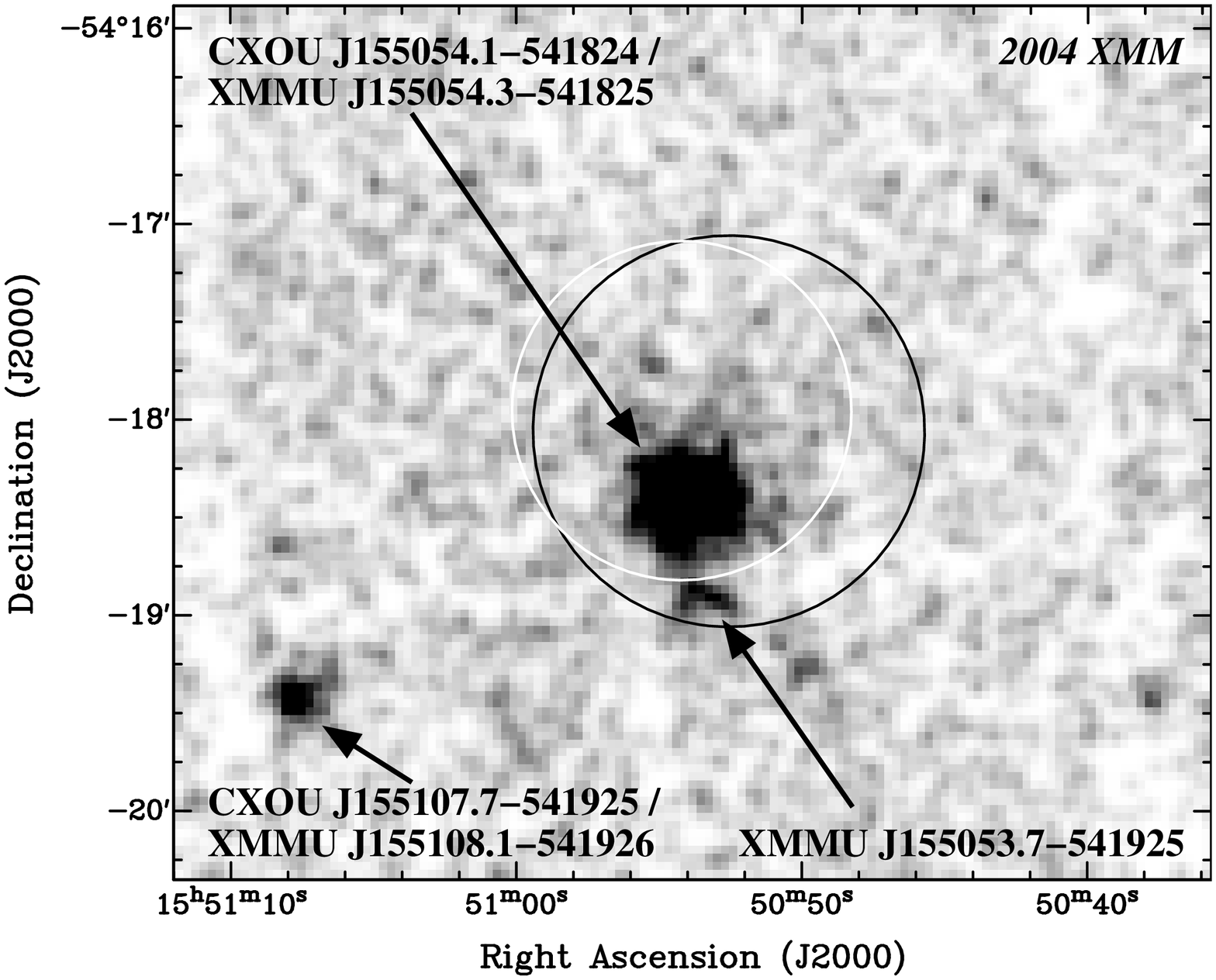}
\includegraphics[angle=0,scale=0.45]{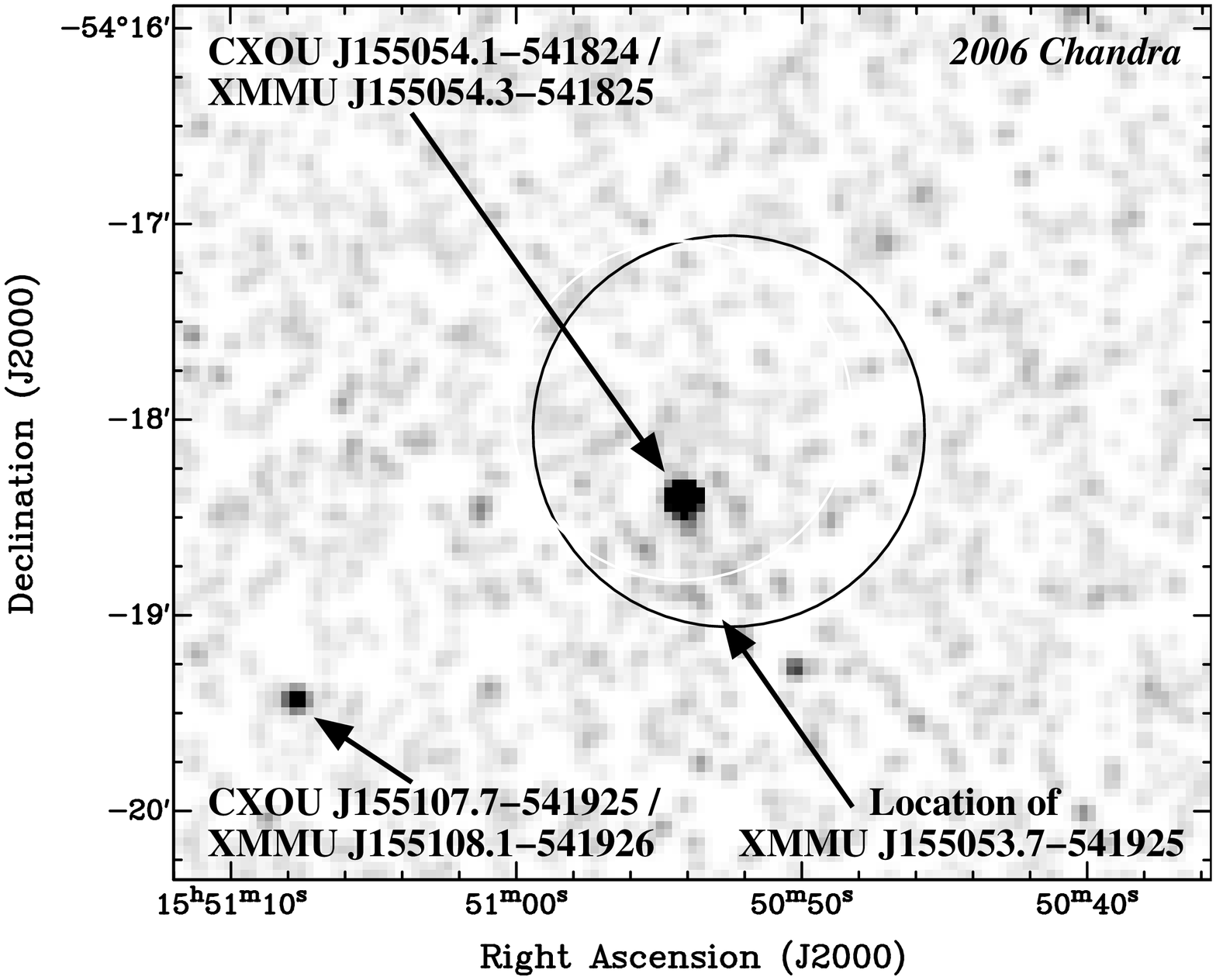}
\end{center}
\figcaption{X-ray images of the field around 1E~1547.0-5408 obtained
  during the 2004 {\it XMM} observation ({\it top}) and 2006 {\it
  Chandra} {\sc Acis--I} observation ({\it bottom}).  The {\it XMM}
  image combines data collected by the {\sc Mos1}, {\sc Mos2}, and
  {\sc pn} instruments.  Both images have been smoothed by a
  5\arcsec~Gaussian.  The white circle indicates the position and
  positional uncertainty of 1E~1547.0-5408 as measured by {\it
  Einstein}, the black circle indicates the position and positional
  uncertainty of 1E~1547.0-5408 and AX~J155052-5418, while the arrows
  indicate the position of
  CXOU~J155054.1-541824=XMMU~J155054.3--541825 as well as unrelated
  field sources XMMU~J155053.7-541925 and
  CXOU~J155107.7--541925=XMMU~J155108.1--541926. \label{xrayimg}}
\end{figure}

As shown in Fig.~\ref{xrayimg}, due to the improved spatial resolution
of {\it XMM} and {\it Chandra}, we detect two point sources in the
overlapping position error circles of 1E~1547.0-5408 and
AX~J155052-5418 -- a bright source that we designate
CXOU~J155054.1-541824 (=XMMU~J155054.3--541825) and a much fainter
source we designate XMMU~J155053.7-541925.  This second source is
detected in both {\it XMM} observations but not the {\it Chandra}
observation because it fell in a chip gap.  For both sources, their
extents are consistent with the point spread function of the telescope
at that position.  The most accurate position for
CXOU~J155054.1-541824 comes from the {\it Chandra} observation, which
gives $\alpha_{\rm J2000}=$15:50:54.11 and $\delta_{\rm
J2000}=$--54:18:23.8.  This position has not been registered to an
external reference frame, due to the lack of field X-ray sources with
counterparts at other wavelengths.  Therefore, the error in this
position is dominated by the pointing accuracy of {\it Chandra} -- a
typical 99\% confidence radius of $\sim0\farcs8$.  As shown in Table
\ref{src2cnt}, the observed count rates of both CXOU~J155054.1-541824
and XMMU~J155053.7-541925 declined between the 2004 and 2006 {\it XMM}
observations -- in the case of CXOU~J155054.1-541824, the count rate
declined by $\sim30\%$, while for XMMU~J155053.7-541925 the count rate
declined by $\sim85\%$.

We have quantified the apparent variation of the X-ray flux from
CXOU~J155054.1-541824 via a more detailed spectral analysis of the
observations listed in Table \ref{obssum}.  Here, and in the
subsequent discussion, we assume that 1E~1547.0-5408 and
AX~J155052-5418 correspond to the same X-ray source, and that this
source is a blend of CXOU~J155054.1-541824, XMMU~J155053.7-541925, and
other adjacent field sources that can be seen at the high angular
resolution of {\it XMM} and {\it Chandra}, but not with {\it ASCA} and
{\it Einstein}.  Since CXOU~J155054.1-541824 substantially dominates
the emission, we here on use 1E~1547.0-5408 to indicate this main
source, except with a distinction with fainter, adjacent sources needs
to be made.  For the {\it Einstein} observation, we used the procedure
defined by \citet{mcgarry05} to determine the flux of 1E~1547.0-5408.
To do this, we assume the X-ray spectrum of 1E~1547.0-5408 is well
modeled by an absorbed power law assuming a hydrogen column density
$N_H=4.2\times10^{22}$~cm$^{-2}$ and a photon index $\Gamma=4.7$, as
derived below from spectral fits to the {\it ASCA}, {\it XMM}, and
{\it Chandra} observations.  To determine the X-ray flux of
1E~1547.0-5408 as measured by {\it ASCA}, {\it XMM}, and {\it
Chandra}, we simultaneously fit the observed spectrum of this source
to an absorbed power law model using {\sc Xspec} v.11.3.1.  We chose
an absorbed power law model since this model provides a good fit to
each of these spectra individually.  For the {\it XMM} and {\it
Chandra} observations, spectral regions were chosen to minimize
contamination from XMMU~J155053.7-541925 and other field sources.  For
the {\it ASCA} observation, this was not possible due to the poor
point spread function of this instrument.  Holding the $N_H$,
$\Gamma$, and normalization constant between these different
observations, which implies a constant flux, resulted in a poor fit
(reduced $\chi^2=2.19$).  Allowing the normalizations to vary
independently, but holding $N_H$ and $\Gamma$ fixed, lowered the
reduced $\chi^2$ to 1.26, a substantial improvement.  As shown in
Table \ref{xrayfluxes}, the results from this analysis imply that
1E~1547.0-5408 is a variable X-ray source.

\begin{figure}
\begin{center}
\includegraphics[angle=0,scale=1.0]{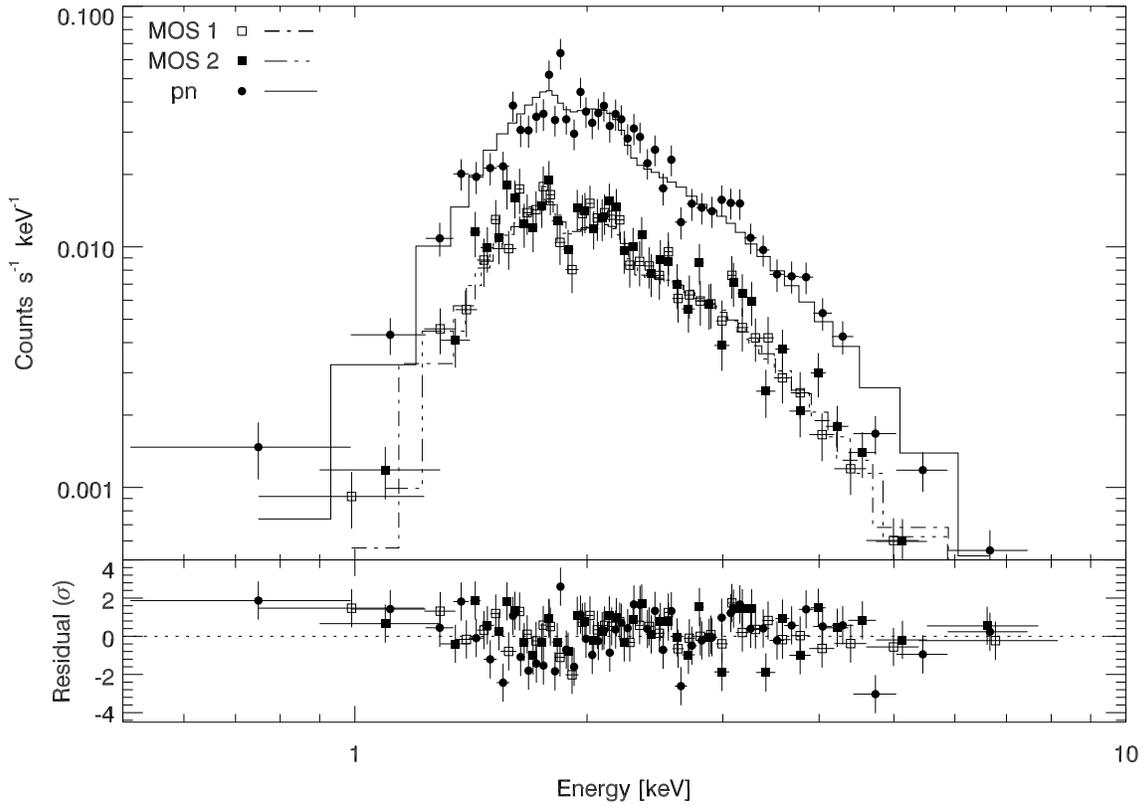}
\end{center}
\figcaption{The X-ray spectrum of CXOU~J155054.1-541824, as measured
  by the {\sc Mos1}, {\sc Mos2}, and {\sc pn} detectors in the 2006
  {\it XMM} observation.  The {\sc Mos1} and {\sc Mos2} channels were
  binned such that there was a minimum of 25 counts per bin, and the
  {\sc pn} channels were binned such that there was a minimum of 50
  counts per bin.  The lines indicates the fit of the absorbed power
  law plus blackbody model given in Table~\ref{xrayspec}, and the
  bottom panel indicates the number of standard deviations the data
  deviates from the model. \label{src1spec}}
\end{figure}

To determine what physical model best describes the X-ray spectrum of
1E~1547.0-5408, we used {\sc Xspec} v.11.3.1 to jointly fit the
spectrum of this source measured during the 2006 August {\it XMM}
observation by the {\sc Mos1}, {\sc Mos2}, and {\sc pn} instruments to
a number of different models, including a power law, bremsstrahlung,
power-law plus blackbody, and two blackbodies, all attenuated by
interstellar absorption.  During this observation, between 0.5 and
10~keV a total of 1005$\pm$33 counts above the background were
collected from this source by the {\sc Mos1} detector, 1062$\pm$34 by
the {\sc Mos2} detector, and 2351$\pm$51 by the {\sc pn} detector.
The resultant X-ray spectrum is shown in Fig. \ref{src1spec}.  As
shown in Table \ref{xrayspec}, an absorbed power-law plus blackbody
model produces the best fit.  According to the f-test, the decrease in
$\chi^2$ for the best two-component model (absorbed power-law plus
blackbody) over the best one-component model (absorbed bremsstrahlung)
is statistically significant at 99.9\% confidence, so the use of
second component is justified.  Therefore, we conclude that either an
absorbed power-law plus blackbody or an absorbed two blackbodies
(these two models produced statistically indistinguishable fits)
provides the best description of the X-ray spectrum of 1E~1547.0-5408.

We also searched for pulsed X-ray emission from 1E~1547.0-5408 using
the data collected during the 2006 {\it XMM} observation using the
$Z_n^2$ test \citep{buccheri83}.  To do so, we first extracted events
from two circular regions centered around the position
CXOU~J155054.1-541824, one $9\farcs5$ in radius and the other
$26\farcs5$ in radius, from the filtered {\sc Mos1}, {\sc Mos2}, and
{\sc pn} datasets.  We then barycentered the arrival times of these
photons to the solar system reference frame.  This search was
conducted on the {\sc Mos1} (0.9s time resolution), {\sc Mos2} (0.9s
time resolution), and {\sc pn} (73.4~ms time resolution) datasets
individually, as well as on them jointly.  To allow for the
possibility that the pulsed fraction might have a strong energy
dependence, we applied various energy cuts to the event lists derived
from both spatial regions.  For all searches the minimum frequency was
$2\times10^{-5}$~Hz, the maximum frequency was 0.6~Hz for the {\sc
Mos} and joint datasets and 6.8~Hz for the {\sc pn} data, and the
frequency step was $2\times10^{-6}$~Hz, oversampling the Nyquist
frequency by a factor of five.  For each combination of spatial
region, detectors, and energy range, we searched for periods summing
up to a maximum harmonic $n=1,2,...,10$.  In none of these different
combinations did we detect a statistically significant signal.
Between 0.5 and 10~keV, the most sensitive dataset for a sinusoidal
($n=1$) pulse profile comes from combining the three detectors and
using the large spatial region.  In this dataset, we have 5305 photons
in 52201 independent trials.  Using the equations in \citet{leahy83},
we are able to place a 5$\sigma$ upper limit on the peak-to-peak pulse
fraction $f_{\rm pulse}$, defined as \citep{patel03}:
\begin{eqnarray}
\label{pulsefrac}
f_{\rm pulse} & = & \frac{N_{\rm max}-N_{\rm min}}{N_{\rm max}+N_{\rm min}}
\end{eqnarray}
where $N_{\rm max}$ and $N_{\rm min}$ are the maximum and minimum
number of counts in the pulse profile, of $f_{\rm pulse}<14\%$.

\subsection{Near-IR Observation of 1E~1547.0-5408}
\label{nir}

\begin{figure}
\begin{center}
\includegraphics[angle=-90,scale=0.75]{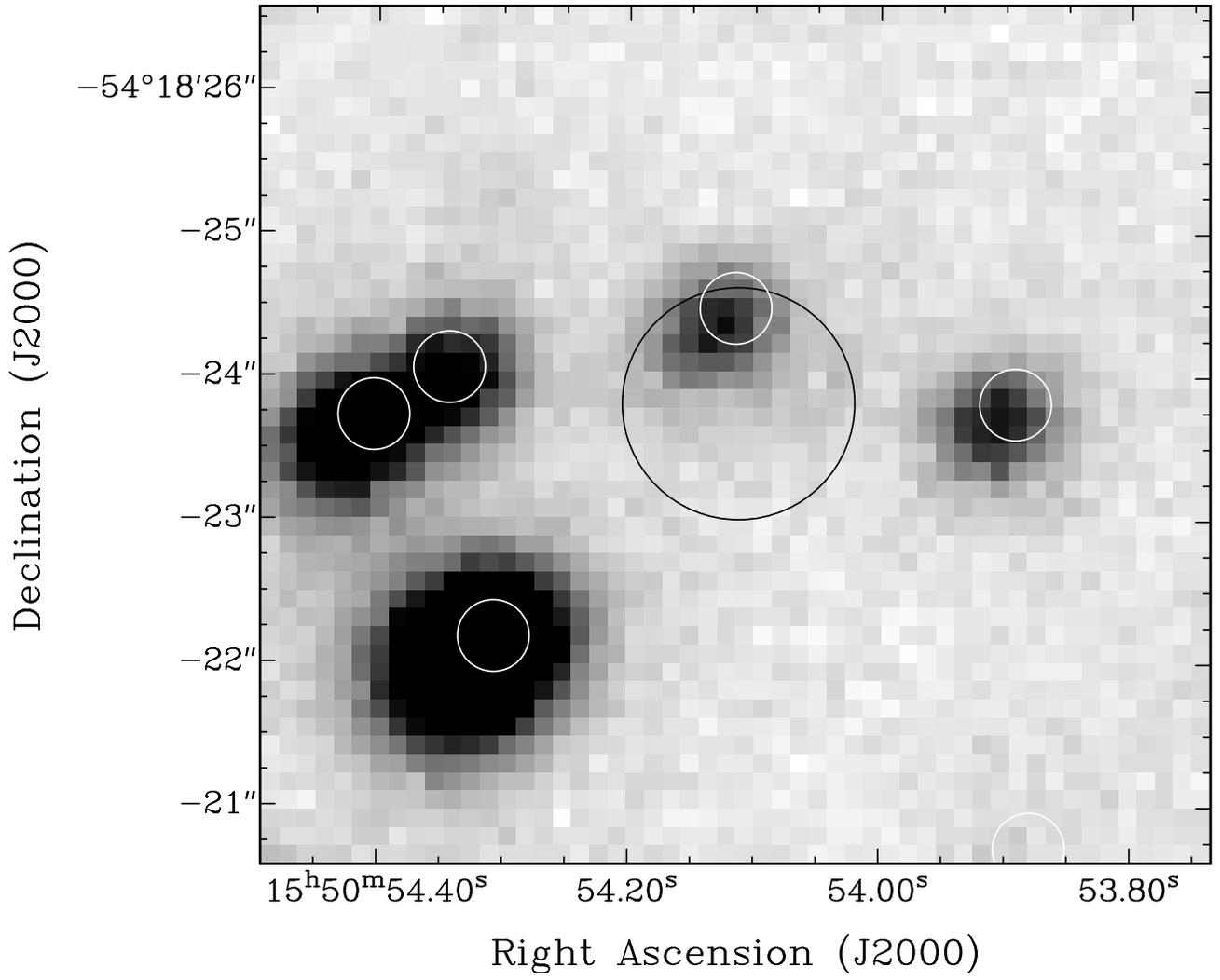}
\end{center}
\figcaption{ $K_s$ image ($\lambda=2.15\mu$m) of the field around
  CXOU~J155054.1-541824 (linear greyscale), overlaid with the 99\%
  positional error circle of CXOU~J155054.1-541824 as measured by {\it
  Chandra} (black circle) and the location of near-IR sources in this
  field as determined by the {\it IRAF} package {\tt daofind} (white
  circles). \label{nirimg}}
\end{figure}

Using the position of CXOU~J155054.1-541824 as measured by {\it
Chandra}, we searched for a near-IR counterpart in a three minute
$K_s$ ($\lambda=2.15\mu$m) observation of this field taken on 13 June
2006 with the Persson's Auxiliary Nasmyth Infrared Camera ({\sc
panic}) instrument on the 6.5m Baade Magellan telescope at the Las
Campanas Observatory in Chile.  This image was sky-subtracted using
the standard procedures in the {\sc iraf} software package, and
registered using the position of 2MASS sources in the field.  We used
the {\sc iraf} task {\tt daofind} to identify objects in this field
and measure their instrumental magnitudes, and used the 144 2MASS
stars in this field with one and only counterpart within 1\arcsec~of
their position to determine the conversion between instrumental and
astronomical magnitudes.  To search for a counterpart, we used the
99\% error ($\sim0\farcs8$) of the {\it Chandra} position quoted in
\S\ref{xray}.  As shown in Fig.~\ref{nirimg}, only one source was
detected inside this region, and this source has an observed magnitude
of $K_s=15.9\pm0.2$.  Using the source magnitude distribution of
objects detected in our $K_s$ observation, we determine that our image
is complete to sources with a $K_s$ magnitude $\la 17.5$.

\section{The Galactic radio shell G327.24-0.13}
\label{radio}

\begin{figure}
\begin{center}
\includegraphics[angle=270,scale=0.7]{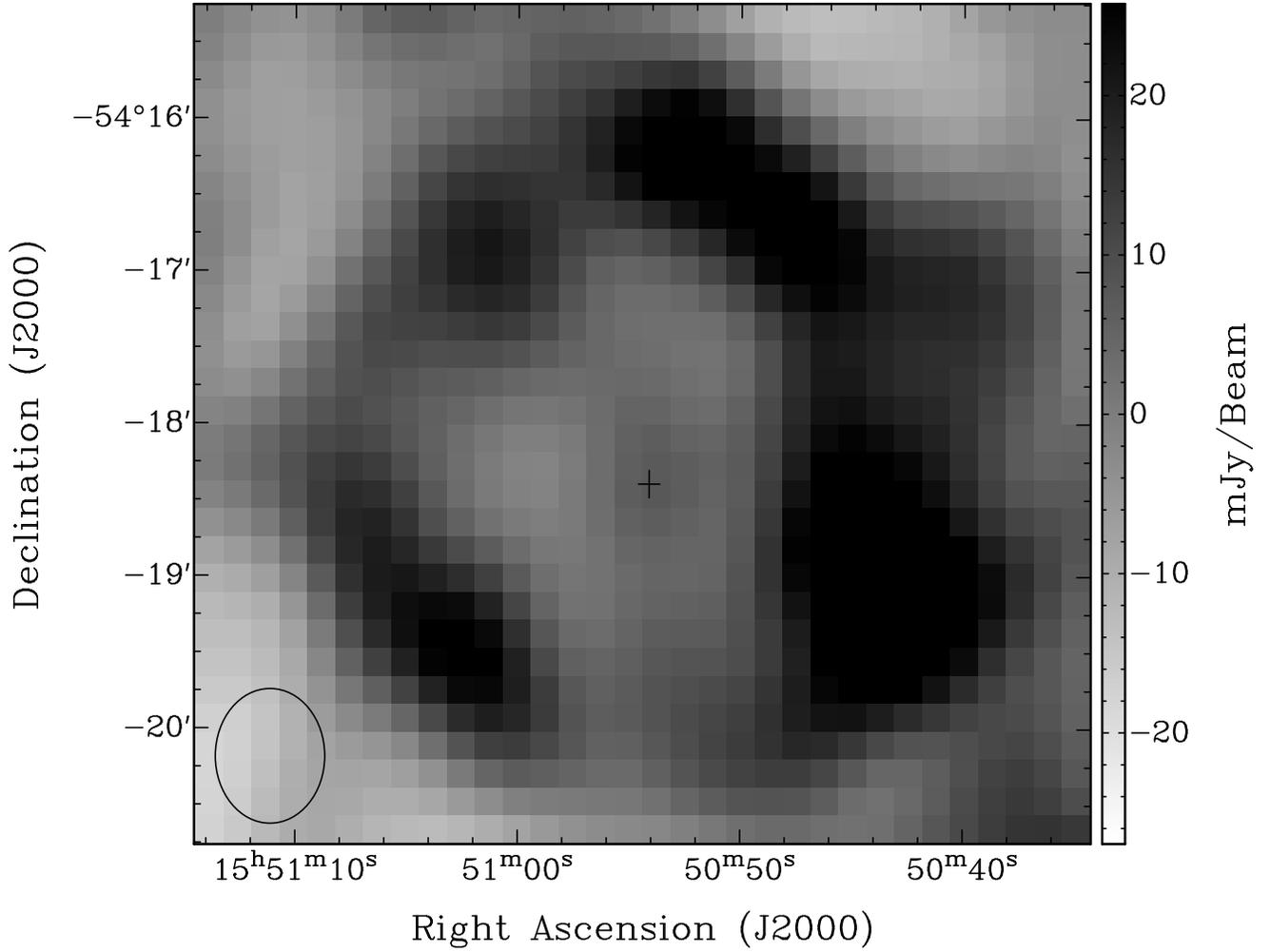}
\end{center}
\figcaption{843~MHz image of G327.24-0.13 from the SUMSS Survey
  \citep{bock99}, overlaid with the {\it Chandra} position of
  CXOU~J155054.1-541824 (black cross; significantly larger than the
  99\% positional uncertainty of $\sim0\farcs8$).  The synthesized
  beam of the radio observation is shown in the lower-left corner
  (black ellipse), and has dimensions
  $53\arcsec\times43\arcsec$. \label{radimg}}
\end{figure}

1E~1547.0-5408 falls within the field of view of multiple recent
southern hemisphere radio surveys: the Molonglo Galactic Plane survey
(MGPS, 843~MHz; \citealt{green99}), the Sydney University Molonglo Sky
Survey (SUMSS, 843~MHz; \citealt{bock99,green02}) and in both the test
region \citep{gaensler01b} and survey region \citep{sgps} of the
Southern Galactic Plane Survey (SGPS, 1.4~GHz).  All four surveys
detected a faint $\sim4^{\prime}$ diameter shell centered on
CXOU~J155054.1-541824, which we designate G327.24-0.13.  In both the
MGPS and SUMSS (Figure~\ref{radimg}) images, there is some evidence
for enhanced emission from the center of G327.24-0.13.  At both
frequencies, the flux was determined by subtracting the observed flux
from G327.24-0.13 by estimates for the diffuse Galactic background at
this position obtained using nearby regions, and the error is
dominated by the uncertainty in the background.  At 843~MHz, the flux
density of G327.24-0.13 -- excluding any interior emission -- is
$0.5\pm0.1$~Jy while at 1.4~GHz, the flux density is $0.3\pm0.1$~Jy.
These fluxes correspond to a radio spectral index between 843~MHz and
1.4~GHz of $\alpha=-0.9\pm0.6$ ($S_{\nu}\propto\nu^\alpha$).

\section{Discussion}
\label{conclusions}

In this section, we use the observational results presented in
\S\ref{1e1547} and \S\ref{radio} to determine the nature of
1E~1547.0-5408 and G327.24-0.13.  Based on the results of the {\it
XMM} and {\it Chandra} observations, we believe 1E~1547.0-5408 is a
blend of compact X-ray sources dominated by the X-ray emission of
CXOU~J155054.1-541824.  Therefore, we conclude that
XMMU~J155053.7-541925 and the other sources detected by {\it XMM}
and/or {\it Chandra} are negligible contributors to the X-ray
properties of 1E~1547.0-5408 as measured by {\it Einstein} and {\it
ASCA}.  We also believe that these sources are unrelated to
CXOU~J155054.1-541824.

Since CXOU~J155054.1-541824 is unresolved by {\it Chandra},
1E~1547.0-5408 is most likely either an active galactic nucleus (AGN),
a neutron star (either a rotation-powered pulsar, a magnetar, or a
compact central object (CCO)), a non-degenerate star, a X-ray binary,
or a cataclysmic variable (CV). A major clue into the nature of
1E~1547.0-5408 is its possible association with a near-IR source, as
discussed in \S\ref{nir}.

\begin{figure}
\begin{center}
\includegraphics[angle=0,scale=1.0]{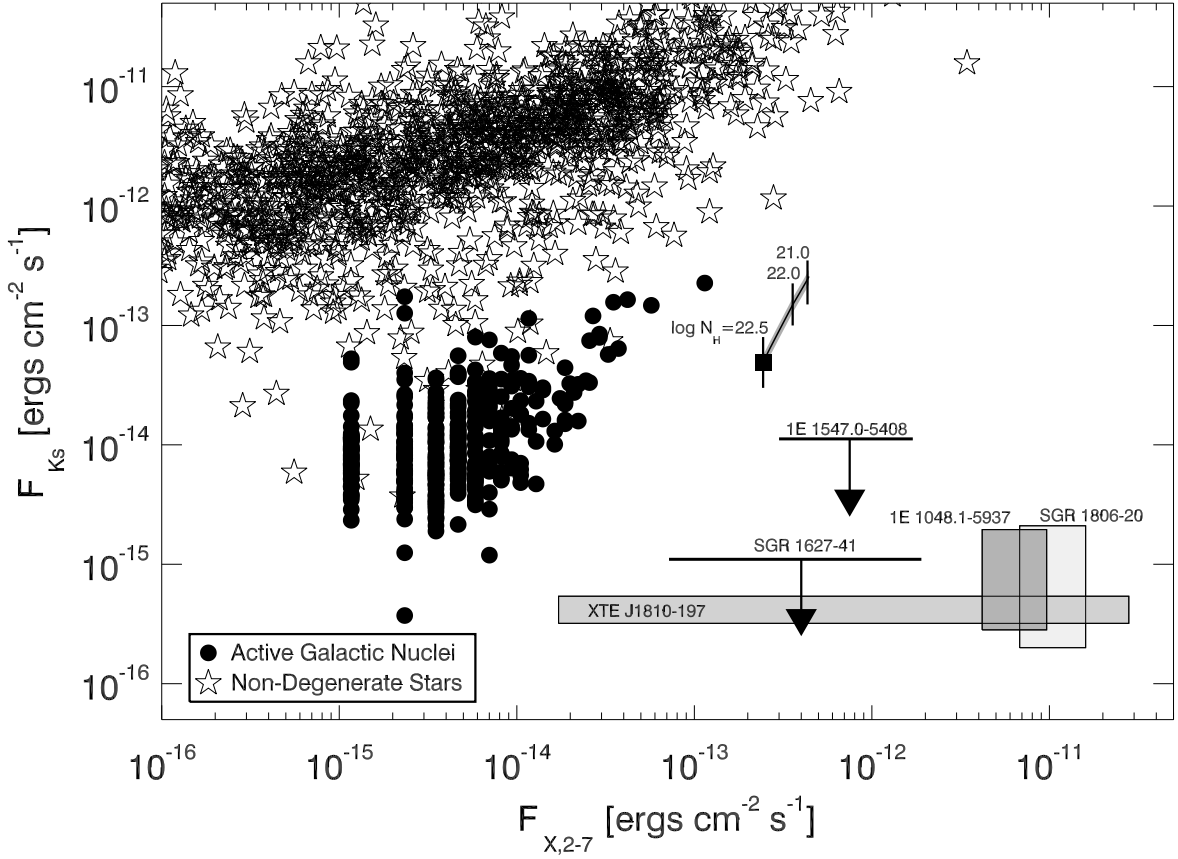}
\end{center}
\figcaption{Observed IR Flux ($F_{K_s}$) versus Absorbed 2--7~keV
  X-ray ($F_{X,2-7}$) flux of 1E~1547.0-5408, sources in the XBo\"otes
  survey (filled circles) believed to be dominated by Active Galactic
  Nuclei, sources in the {\it Chandra} Orion Ultradeep Project (open
  stars) believed to be non-degenerate stars, and selected magnetars
  -- both magnetars with detected near-IR counterparts (shaded
  rectangles) and without (upper limits). If 1E~1547.0-5408 is
  associated with the near-IR source within its positional error
  circle shown in Figure \ref{nirimg}, the observed location of
  1E~1547.0-5408 is the filled square, while the locations of
  1E~1547.0-5408 for different values of $N_H$ are shown by the black
  line, with the gray rectangle denoting the 1$\sigma$ error in
  $F_{K_s}$.  The units of $N_H$ are cm$^{-2}$, and the location of
  1E~1547.0-5408 for $0<N_H<10^{21}$~cm$^{-2}$ is indistinguishable
  from the location for $N_H=10^{21}$~cm$^{-2}$ on this plot. The
  sources in the XBo\"otes survey are gridded into columns due to the
  small number of X-ray photons detected from most of these sources.
  The X-ray and IR data for the magnetars, along with those for
  1E~1547.0-5408 (assuming that it is not detected in our near-IR
  observation, and thus only have an upper limit on its near-IR
  flux), are given in Table \ref{xrayirdata}.
  \label{fluxratimg}}
\end{figure}

If 1E~1547.0-5408 is associated with this near-IR source, it is most
likely either a non-degenerate star in the Milky Way or a AGN.  If a
non-degenerate star, then the absorbed bremsstrahlung is the most
realistic description of the X-ray emission.  If an AGN, only the
absorbed bremsstrahlung fit to the observed spectrum gives parameters
similar to that observed from other AGN.  To determine if either
identification is reasonable, we compared the X-ray and IR fluxes of
this source to those of known stars and AGN, similar to the approach
used by \citet{kaplan04}.  For stars, we used data obtained by the
{\it Chandra} Orion Ultradeep Project \citep{getman05}, and for AGN,
we used the X-ray \citep{kenter05} and near-IR \citep{jannuzi04} data
from the XBo\"{o}tes survey.  For both surveys, we only used the
observed 2--7~keV flux since this quantity is less sensitive to
interstellar absorption and choice of X-ray spectral model than the
0.5--2~keV flux.  To correct for the difference in $N_H$ observed
towards 1E~1547.0-5408 from the Galactic value of $N_H$ towards
sources in the Bo\"otes field ($N_H=10^{20}$~cm$^{-2}$;
\citeauthor{kenter05} \citeyear{kenter05}) and the value observed
towards sources in the Orion field ($N_H = 10^{21}-10^{23}$~cm$^{-2}$;
\citeauthor{getman05} \citeyear{getman05}), we determined the 2--7~keV
X-ray and near-IR fluxes of 1E~1547.0-5408 over this range of $N_H$.
As shown in Figure \ref{fluxratimg}, the IR and X-ray properties of
1E~1547.0-5408 are inconsistent with both populations.  Therefore, we
conclude that 1E~1547.0-5408 is not associated with this near-IR
source.  To compute the probability of their association being a
coincidence, we shifted the RA and DEC of 1E~1547.0-5408 by a random
amount between $\pm0\farcs8-10\arcsec$, and determined if there is a
near-IR source within $0\farcs8$ of its adjusted position.  This
analysis implies a false coincidence rate of $\sim$14\%, implying that
there is a reasonable possibility that the near-IR object is an
unrelated field source.  As a result, we adopt a $K_s$ magnitude of
17.5 as an upper limit on $K_s$ magnitude 1E~1547.0--5408.

As a result, we are left with the possibility that this X-ray source
is a neutron star, X-ray binary, or a CV.  The location of
1E~1547.0-5408 in the center of SNR candidate G327.24-0.13 strongly
implies that this source is a neutron star.  This identification is
supported by the fact that the bremsstrahlung and power-law fits to
the X-ray spectrum of 1E~1547.0-5408 shown in Table \ref{xrayspec} are
inconsistent with the spectra expected from X-ray binaries (a modified
blackbody with kT$\sim$1-2~keV; \citeauthor{white88}
\citeyear{white88}) and CVs (bremsstrahlung emission with kT$>1$~keV;
\citeauthor{eracleous91} \citeyear{eracleous91}).  The X-ray spectrum
of 1E~1547.0-5408 is also inconsistent with that observed from
rotation-powered pulsars (a power law with $\Gamma=1-2$;
\citeauthor{cheng04} \citeyear{cheng04}).  However, the parameters for
the absorbed power law, the absorbed power law and blackbody, and the
absorbed two blackbody models are similar to that observed from both
magnetars \citep{mereghetti01} and CCOs \citep{pavlov04}.  As shown in
Table \ref{xrayirdata}, the observed X-ray variability of
1E~1547.0-5408 described in \S\ref{xray} is similar to that observed
from several magnetars. CCOs, on the other hand, appear to be steady
X-ray sources, and therefore we conclude the 1E~1547.0-5408 is most
likely a magnetar.  One crucial difficulty with this interpretation is
our failure to detect pulsed X-ray emission from this source.
Magnetars typically have high pulsed fractions in the X-ray regime,
for example before its recent outburst CXOU~J164710.2-455216 had a
pulsed fraction of $\sim50\%$ \citep{muno06}, much higher than the
5$\sigma$ upper limit derived in \S\ref{xray} of 14\% on the 0.5--10~keV
peak-to-peak pulsed fraction from 1E~1547.0-5408 for a sinusoidal
pulse profile, the characteristic pulse profile for magnetars.
However, this upper limit is higher than the pulsed fraction of at
least one magnetar, 4U~0142+61, which has a 0.5--7~keV peak-to-peak
pulsed fraction of $\approx7\%$ \citep{patel03,gohler05}.

If this identification is correct, we can compare the X-ray properties
of 1E~1547.0-5408 to those of other magnetars.  For the power-law plus
blackbody model of the spectrum, the 2--10 keV unabsorbed flux of
1E~1547.0-5408 ranges from
$\sim(0.5-3)\times10^{-12}$~ergs~cm$^{-2}$~s$^{-1}$.  For a distance
$d=4d_4$~kpc (it will be argued below that $d_4 \approx 1$ is a
reasonable distance estimate), this translates to a luminosity range
of $\sim(0.9-7)d_4\times10^{33}$~ergs~s$^{-1}$.  Comparing this
value to those given in the SGR/AXP Online Catalog\footnote{Available
online at {\tt
http://www.physics.mcgill.ca/$\sim$pulsar/magnetar/main.html}.}, this
falls within the X-ray luminosity range spanned by confirmed
magnetars.  With this information, we can also compare the near-IR and
X-ray properties of 1E~1547.0-5408 to that of other magnetars, whose
properties are listed in Table \ref{xrayirdata}.  As shown in Figure
\ref{fluxratimg}, the range of X-ray fluxes observed from
1E~1547.0--5408, as well as the upper limit on the near-IR flux is
consistent with that observed from confirmed magnetars.

The radio spectral index of G327.24-0.13 derived in \S\ref{radio} is
consistent with a non-thermal origin, implying that this source is
either a stellar wind bubble or a SNR.  The lack of any mid-IR
counterpart to G327.24-0.13 in the {\it Spitzer} {\sc glimpse} survey
\citep{glimpse} gives support to the latter interpretation.  If
1E~1547.0-5408 is a magnetar and G327.24-0.13 is the SNR created by
its progenitor, then this system becomes a new addition to the handful
of associations between magnetars and SNRs. Assuming that the two
sources are indeed associated, we can obtain initial constraints on
the possible distance to 1E~1547.0--5408/G327.24-0.13 as follows.
Recent evidence has suggested that magnetar progenitors are especially
massive stars \citep{gaensler05, muno06}, and therefore expect
significant star--formation activity in their vicinity.  Indeed, in
this case there are two nearby ($\la0\fdg5$ away) thermal radio
sources, G326.96+0.03 and G327.99-0.09, both associated with a large
star-forming complex in the Scutum-Crux spiral arm.  H\,{\sc i}\
absorption and hydrogen recombination line measurements indicate that
the distance to G326.96+0.03 and G327.99-0.09 both fall within the
range $d=3.7-4.3$~kpc \citep{caswell87,sgpstest}.  By associating
1E~1547.0-5408/G327.24-0.13 with this region, we argue that $d_4
\approx 1$.  If so, the observed size of G327.24-0.13 ($\sim4\arcmin$)
implies a diameter of $\sim5d_4$~pc, making G327.24-0.13 one of the
smallest, and therefore probably youngest, known SNRs.

As mentioned in \S\ref{xray}, 1E~1547.0-5408 was originally discovered
in a search for the X-ray counterpart of unidentified $\gamma$-ray
source 1CG327--0.  Quiescent emission from magnetars has been detected
at energies as high as $\sim$100~keV by the {\it Integral} satellite
\citep{gotz06}, but not in the $>100$~MeV range detected by {\it
Cos-B} from 1CG327--0 \citep{hermsen77}.  Though magnetars have been
known to be the source of intense $\gamma$-ray flares
\citep{hurley05}, these events are rare (only three have been detected
in the past 30 years), and the maximum photon energy detected from
these flares ($\sim1$~MeV; \citeauthor{hurley05} \citeyear{hurley05})
is significantly less than the photon energies detected from
1CG327--0.  As a result, we do not believe that 1E~1547.0-5408 is the
X-ray counterpart of 1CG327--0.  Another possibility is that 1CG327--0
is associated with G327.24-0.13, since high-energy $\gamma$-ray
emission has been detected from some SNRs.  However, the non-detection
of 1CG327--0 by {\it EGRET} implies this source is variable, making
any association of 1CG327--0 with the SNR candidate G327.24--0.13
unlikely.  Future {\it GLAST} observations of this region should
determine if 1CG327--0 is a real $\gamma$-ray source and, if so, if it
is affiliated with 1E~1547.0-5408 or G327.24-0.13.

To summarize, in this paper we presented X-ray, near-IR, and radio
observations of 1E~1547.0-5408 and of the field around it.  A
consistent explanation of these observation is that 1E~1547.0-5408 is
a magnetar possibly associated with SNR candidate
G327.24--0.13. Deeper X-ray observations of 1E~1547.0-5408 are needed
to confirm its identification as a magnetar by detecting X-ray
pulsations from this source, deep near-IR observations of this source
are needed to discover its near-IR counterpart, and additional radio
observations of G327.24--0.13 are required to determine if it is a
SNR.  The identification of 1E~1547.0-5408 as a magnetar candidate
illustrates the importance of follow-up observations of other bright,
unidentified X-ray sources in the Galactic Plane for understanding the
X-ray population of the Milky Way and for discovering new members of
exotic classes of neutron stars.

\acknowledgements J.D.G and B.M.G. are supported by NASA through LTSA
grant NAG5-13032.  B.M.G. is also supported by an Alfred P. Sloan
Research Fellowship.  J.D.G. would like to thank Julia Bryant for
generously carrying out the Magellan observation as well as the
initial IR data reduction, Ingyin Zaw, Jenny Greene, Craig Heinke,
Wynn Ho, Jae Sub Hong, Xavier Koenig, Dave Monet, Dan Padnaude, and
Pat Slane for useful discussions, Ryan Hickox for providing the X-ray
and IR emission of sources in the Bo\"otes survey, and David Kaplan,
Manual Torres, and Maryam Modjaz for help analyzing the Magellan data.
The Australia Telescope is funded by the Commonwealth of Australia for
operation as a National Facility managed by CSIRO.  The MOST is
operated by the University of Sydney with support from the Australian
Research Council and the Science Foundation for Physics within the
University of Sydney.  This research has used the resources of the
High Energy Astrophysics Science Archive Research Center (HEASARC).
The Two Micron All Sky Survey is a joint project of the University of
Massachusetts and the Infrared Processing and Analysis
Center/California Institute of Technology.

\bibliography{ms}
\bibliographystyle{apj}

\clearpage

\begin{table}
\begin{center}
\scriptsize
\caption{Properties of the new and archival X--ray observations of
  1E~1547.0-5408 presented in this paper \label{obssum}}
\begin{tabular}{cccc}
\hline
\hline
{\sc Date} & {\sc Telescope} & {\sc Instruments} & {\sc Exposure Time} [ks] \\ 
\hline
1980 March 12 & {\it Einstein} & HRI & 3.6 \\
1998 February 24 & {\it ASCA} & GIS & 6.7 \\
2004 February 8 & {\it XMM--Newton} & {\sc Mos1}, {\sc Mos2}, {\sc
  pn} & 8.7, 8.7, 6.2 \\
2006 July 1 & {\it Chandra} & {\sc Acis-I} & 9.5 \\
2006 August 22 & {\it XMM--Newton} & {\sc Mos1}, {\sc Mos2}, {\sc pn}
& 45.1, 45.1, 38.6 \\ 
\hline
\hline
\end{tabular}
\tablecomments{For the 2004 February 8 {\it XMM} observations, the
  {\sc Mos1} detector was operated in Large Window Mode using the
  Medium filter, the {\sc Mos2} detector was operated in Full Window
  Mode using the Thin filter, and the {\sc pn} detector was operated in
  Full Window Mode using the Medium filter.  For 2006 July 1 {\it
  Chandra} observation, the {\sc Acis-I} instrument was operated in
  Faint mode.  For the 2006 August 22 {\it XMM} observation, the {\sc
  Mos1} and {\sc Mos2} detectors were operated in Small Window Mode,
  and the {\sc pn} detector in Full Window Mode, all using the Medium
  Filter.  The definitions of the Window Modes on {\it XMM} can be
  found at {\tt
  http://xmm.esac.esa.int/external/xmm\_user\_support/documentation/uhb/node28.html},
  and the telemetry modes of {\it Chandra} are defined in {\tt
  http://cxc.harvard.edu/proposer/POG/html/ACIS.html\#tth\_sEc6.13.1}.
  For all observations, the quoted exposure time accounts for
  dead-time and the removal of background flares.}
\end{center}
\end{table}

\begin{table}
\begin{center}
\caption{0.5--10~keV background subtracted count rate of
  CXOU~J155054.1-541824 and XMMU~J155053.7-54192 measured during the
  2004 and 2006 {\it XMM} observations \label{src2cnt} }
\begin{tabular}{ccccc}
\hline \hline & \multicolumn{2}{c}{CXOU~J155054.1-541824} &
\multicolumn{2}{c}{XMMU~J155053.7-541925} \\ 
\hline 
{\it Detector} & {\it 2004 February} & {\it 2006 August} & 
{\it 2004 February} & {\it 2006 August} \\ 
\hline 
{\sc Mos1} (counts ks$^{-1}$) & $33\pm2$ & $22.7\pm0.8$ & $4.2\pm0.8$
& $0.5\pm0.2$ \\ 
{\sc Mos2} (counts ks$^{-1}$) & $29\pm2$ & $23.7\pm0.8$ & $2.6\pm0.7$
& $0.4\pm0.2$ \\ 
{\sc pn} (counts ks$^{-1}$) & $85\pm4$ & $62\pm1$ & $\cdots$ & $2.0\pm0.3$ \\ 
\hline 
\hline
\end{tabular}
\end{center}
\tablecomments{The same source and background regions were used for
  both observations.  In the 2004 {\it XMM} observation,
  XMMU~J155053.7-541925 fell in a chip-gap in the {\sc pn} detector
  which is why no count-rate is given for this detector.  Errors
  denote the 1$\sigma$ confidence interval.}
\end{table}

\begin{table}
\begin{center}
\caption{The absorbed 0.5--10~keV flux of 1E~1547.0-5408 as
  measured by the observations listed in Table \ref{obssum}
  \label{xrayfluxes}}
\begin{tabular}{ccc}
\hline
\hline
{\sc Date} & {\sc Observatory} & {\sc Absorbed Flux} [$\times 10^{-12}$ ergs 
  cm$^{-2}$ s$^{-1}$] \\
\hline
1980 March 12 & {\it Einstein} & $1.9_{-0.5}^{+1.0}$ \\ 
1998 February 24 & {\it ASCA} & $2.1^{+0.3}_{-0.3}$ \\ 
2004 February 8 & {\it XMM} & $0.45_{-0.03}^{+0.02}$ \\ 
2006 July 1 & {\it Chandra} & $0.30_{-0.04}^{+0.04}$ \\ 
2006 August 22 & {\it XMM} & $0.31_{-0.03}^{+0.01}$ \\ 
\hline
\hline
\end{tabular}
\end{center}
\tablecomments{The quoted errors denote the 90\% confidence interval.
  The {\it ASCA}, {\it XMM}, and {\it Chandra} fluxes are from a joint
  absorbed power law fit to the spectra, where the hydrogen column
  density ($N_H$) and photon index ($\Gamma$) were held fixed for
  these observations, but the normalizations were allowed to vary.
  This fit produced $N_H=4.2\pm0.2\times10^{22}$cm$^{-2}$ and
  $\Gamma=4.7\pm-0.2$ (errors denote 90\% confidence level of the
  parameters), with a reduced $\chi^2=1.26$ for 263 degrees of
  freedom.  For the {\it Einstein} flux, we determined the 90\%
  interval in the observed count rate, and then converted this count
  rate to a flux using the same method described by
  \citet{mcgarry05}.}
\end{table}

\begin{table}
\begin{center}
\caption{Spectral Fits to 2006 August {\it XMM} observation of
  CXOU~J155054.1-541824 \label{xrayspec}}
\footnotesize
\begin{tabular}{ccccc}
\hline
\hline
{\sc Parameter} & \multicolumn{4}{c}{\sc Value} \\
\hline
\hline
Model & {\bf phabs * pow} & {\bf phabs * bremss} & {\bf phabs *
  (pow+bb)} & {\bf phabs*(bb+bb)} \\ 
\hline
N$_{H}$ [cm$^{-2}$] & $4.3^{+0.3}_{-0.2}\times10^{22}$ &
$3.0\pm0.2\times10^{22}$ & $3.1^{+0.7}_{-0.8}\times10^{22}$ &
$2.5_{-0.2}^{+0.3}\times10^{22}$  \\ 
$\Gamma_1$/$kT_1 [kT] $ & $4.8\pm0.2$ & $1.1\pm0.1$ &
$3.7^{+0.8}_{-2.0}$/$0.43^{+0.03}_{-0.04}$ & $0.45_{-0.06}^{+0.04}$ \\  
$kT_2$ [kT] & $\cdots$ & $\cdots$ & $\cdots$ & $1.1^{+1.4}_{-0.3}$ \\
Absorbed Flux [ergs~cm$^{-2}$~s$^{-1}$] & $3.1_{-0.3}^{+0.1}\times10^{-13}$ &
$3.0_{-0.6}^{+0.3}\times10^{-13}$ & $3.1_{-1.2}^{+0.3}\times10^{-13}$
& $3.1_{-0.9}^{+0.1}\times10^{-13}$ \\   
Unabsorbed Flux [ergs~cm$^{-2}$~s$^{-1}$] & 
$2.8_{-1.9}^{+0.1}\times10^{-11}$ & $2.3_{-0.5}^{+0.1}\times10^{-12}$
& $3.6_{-2.4}^{+0.4} \times10^{-12}$ & $1.0_{-0.2}^{+0.01}\times10^{-12}$ \\ 
$\chi^2$/d.o.f (reduced $\chi^2$) & $147.9/126$ (1.17) & $144.7/126$
(1.14) & $134.4/124$ (1.08) & $135.3/124$ (1.09) \\  
\hline
\hline
\end{tabular}
\end{center}
\tablecomments{The results from an absorbed power-law ({\bf phabs *
  pow}), an absorbed bremsstrahlung ({\bf phabs * bremss}), an
  absorbed power-law plus blackbody ({\bf phabs * (pow+bb)}), and an
  absorbed blackbody plus blackbody ({\bf phabs * (bb+bb)}) fit to the
  {\sc Mos1}, {\sc Mos2}, and {\sc pn} spectra of
  CXOU~J155054.1-541824 obtained during the 2006 {\it XMM}
  observation.  The fits were done between 0.5--10~keV.  The {\sc
  Mos1} and {\sc Mos2} channels were binned such that there were a
  minimum of 25 counts per bin, and the {\sc pn} channels were binned
  such that there were a minimum of 50 counts per bin.  Errors
  indicate 90\% confidence intervals for each quantity, and ``d.o.f''
  stands for degrees of freedom. Both the absorbed and unabsorbed flux
  were calculated between 0.5 and 10~keV.}
\end{table}

\begin{table}
\begin{center}
\tiny
\caption{The 2--7~keV Absorbed X-ray Flux and $K_s$ Flux of
  1E~1547.0-5408 and Some Magnetars \label{xrayirdata}}
\begin{tabular}{cccc}
\hline
\hline
Source & X-ray Flux [ergs cm$^{-2}$ s$^{-1}$] & $K_s$ Flux [ergs
  cm$^{-2}$ s$^{-1}$] & References \\
\hline
1E~1547.0-5408 & $(0.3-1.7)\times10^{-12}$ & $\la1\times10^{-14}$ & $\cdots$ \\
\hline
SGR 1900+14 & $(0.4-1.9)\times10^{-11}$ & $\la5\times10^{-16}$ &
\citet{kaplan02,esposito07}\\ 
SGR 1627-41 & $(0.1-1.9)\times10^{-12}$ & $\la1\times10^{-15}$ &
\citet{mereghetti06,wachter04}\\ 
SGR 1806-20 & $(0.7-1.6)\times10^{-11}$ & $(0.2-2.1)\times10^{-15}$ &
\citet{rea05b,woods07} \\ 
            & & & \citet{kosugi05,israel05} \\
SGR 0526-66 & $(4.1-4.4)\times10^{-13}$ & $\la3\times10^{-16}$ &
\citet{klose04,kulkarni03}\\
XTE J1810-197 & $(0.002-2.8)\times10^{-11}$ & $(3.2-5.4)\times10^{-16}$ &
\citet{halpern05,gotthelf05}\\ 
              & & & \citet{rea04} \\
1E 1048.1-5937 & $(4.2-9.7)\times10^{-12}$ & $(0.3-2.0)\times10^{-15}$ &
\citet{tiengo05} \\ 
               & & & \citet{wang02,durant05} \\
4U 0142+61 & $(5.6-6.7)\times10^{-11}$ & $(0.6-1.5)\times10^{-15}$ &
           \citet{gohler05,patel03} \\ 
           & & & \citet{durant06} \\
CXOU J164710.2-455216 & $(0.001-1.7)\times10^{-11}$ &
           $\la4\times10^{-16}$ & \citet{muno07,wang06} \\ 
1RXS J170849.0-400910 & $(2.6-2.8)\times10^{-11}$ & $(2.2-3.2)\times10^{-15}$ &
           \citet{campana07,durant06b} \\ 
\hline
\hline
\end{tabular}
\end{center}
\tablecomments{Only confirmed magnetars which had sufficient X-ray
  spectral information available in the literature are included in
  this table.  The X-ray fluxes quoted in this table are the absorbed
  flux, and no corrections have been made to either the X-ray flux or
  $K_s$ fluxes for the different values of $N_H$ measured towards
  these sources.  The X-ray flux range of 1E~1547.0-5408 was
  calculated using the same procedure described in the caption of
  Table \ref{xrayfluxes}.}
\end{table}

\clearpage

\end{document}